# On the Asymptotics of the Finite–Perimeter Partition Function of Two–Dimensional Lattice Vesicles

T. Prellberg‡* and A. L. Owczarek§†

‡Department of Theoretical Physics, University of Manchester

Manchester M13 9PL, United Kingdom

§Department of Mathematics, University of Melbourne,

Parkville, 3052, Australia

September 10, 1998

**Abstract**

We derive the dominant asymptotic form and the order of the correction terms of the finite–perimeter partition function of self–avoiding polygons on the square lattice, Lwhich are weighted according to their area $A$ as $q^A$, in the inflated regime, $q > 1$. The approach $q \to 1^+$ of the asymptotic form is examined.

*email: `prel@a13.ph.man.ac.uk`

†email: `aleks@maths.mu.oz.au`

# 1 Introduction

A simple model of a closed, fluctuating membrane in solution (or vesicle), such as those found in biological contexts, is a self–avoiding surface on a $d$–dimensional hypercubic lattice. To take account of the effects of factors such as osmotic pressure and pH differences between the inside and outside of the membrane it is advantageous to sort the configurations according to their volume and surface area. In two dimensions, self–avoiding polygons (SAP) weighted by area and perimeter were investigated by Fisher *et al.* [9] after the general problem of two–dimensional vesicles was discussed by Leibler *et al.* [13]. Exact enumerations of SAP by area and perimeter, and some related rigorous results on the mean area of polygons of fixed perimeter have also been given [12, 8], after pioneering work of Hiley and Sykes [11] on their enumeration.

A vesicle in two dimensions will be modelled in this paper by a self–avoiding polygon on the square lattice, where both the perimeter and area are controlled in some fashion. To be more precise, one quantity often considered when investigating the behaviour of lattice vesicles is the finite–perimeter partition function. This is defined as

$$Z_n(q) = \sum_m c_m^n q^m , \qquad (1.1)$$

where $c_m^n$ is the number of some set of polygon configurations enumerated with respect to their perimeter, $2n$, and area, $m$, and the sum is over all possible values of $m$. (Since only the square lattice is considered here, where the perimeter of the polygons contains an even number of bonds, we will use the convention that $n$ denotes *half* of the length of the perimeter.) It is this quantity that will be the focus of our work here, more precisely, its asymptotic behaviour as $n \to \infty$ for a fixed value of $q$. Moreover, everywhere we will restrict $q$ to be larger than one, that is, $q > 1$. In the course of our discussion we will consider several subsets of self–avoiding polygons on the square lattice: these include convex polygons, directed convex polygons, Ferrers diagrams and simple rectangles. The general area–perimeter counting problem for these subsets have been examined previously [4, 5, 7, 1, 2, 3, 6, 14, 15, 16, 17]. In particular, the definitions, including diagrams, of the various polygon models can be found in Bousquet–Mélou [2]. However, their finite–perimeter partition functions' asymptotics for $q > 1$ have not been explicitly examined.

In this paper we prove that in two dimensions for SAP

$$Z_n(q) = A(q)\, q^{n^2/4} \left(1 + O(\rho^n)\right) \text{ as } n \to \infty \; , \tag{1.2}$$

for some $0 < \rho < 1$, where $A(q) = A_o(q)$ or $A(q) = A_e(q)$ when $n$ is restricted to subsequences with $n$ being odd or even respectively. We give *explicit* expressions for $A_o(q)$ and $A_e(q)$. In fact we show that these functions coincide with those obtained if one only considered *convex* polygons. Note also that the odd/even dichotomy implies there is not a unique asymptotic form for $Z_n(q)$ in the regime $q > 1$.

We also deduce that there is an essential singularity in both the $A(q)$ functions as $q$ approaches 1 from above; in particular

$$A(q) \sim \frac{1}{4}\left(\frac{\varepsilon}{\pi}\right)^{3/2} e^{2\pi^2/3\varepsilon} \qquad \text{as } \varepsilon = \log q \to 0^+ \tag{1.3}$$

for both even and odd $n$.

In Fisher *et al.* [9] there is an argument giving the leading order factor of the finite–perimeter partition function asymptotics for polygons. The partition function $Z_n(q)$ is bounded for $q > 1$ by

$$q^{M(n)} \leq Z_n(q) \leq q^{M(n)} Z_n(1) = q^{M(n)} \mu_{saw}^{2n+o(n)} \tag{1.4}$$

where $M(n)$ is the maximal area of a polygon with perimeter $2n$ and $\mu_{saw}$ is the connectivity constant for self–avoiding walks. From this and the exact value of $M(n)$ (see 3.1) it follows immediately that

$$Z_n(q) = q^{n^2/4} e^{O(n)} \text{ as } n \to \infty \; . \tag{1.5}$$

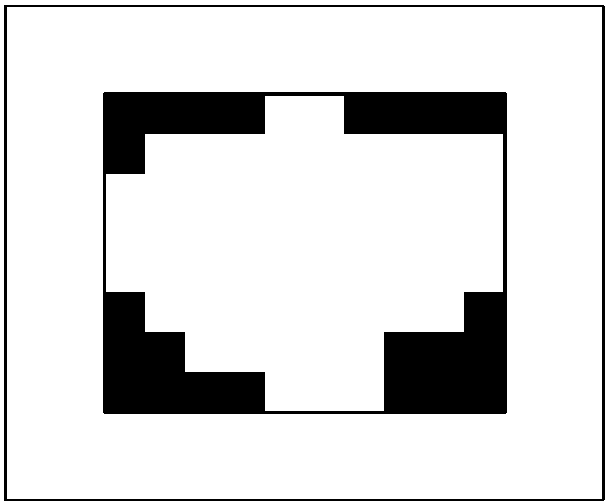

Figure 1: Pictorial representation of (1.6): the partition function is asymptotically dominated by convex polygons, which are constructed from rectangles by removing corners made of Ferrers diagrams.

To refine this result, we show that in fact for all $q > 1$ the partition function asymptotics is completely dominated by the convex configurations. This is stated in Theorem 2.1. In Theorem 2.2 we then discuss the asymptotics for various models of convex polygons. Taken together these two theorems enable the following explicit expression, described precisely in Corollary 2.3, for the leading asymptotic behaviour of $Z_n(q)$ to be given

$$Z_n(q) = \frac{(1 + O(\rho^n))}{(q^{-1}; q^{-1})_\infty^4} \sum_{k=-\infty}^{\infty} q^{k(n-k)} \tag{1.6}$$

for some $0 < \rho < 1$. Here,

$$(x; q)_m \stackrel{def}{=} \prod_{k=1}^{m} (1 - xq^{k-1}) \tag{1.7}$$

is the standard $q$–product notation. This is the main result of our work.

The asymptotic form (1.6) has a straightforward combinatorial interpretation (see Figure 1). The infinite sum has its origin in the generating function for rectangles

$$\sum_{k=1}^{n-1} q^{k(n-k)} \ . \tag{1.8}$$

(A rectangle of perimeter $2n$ may have sides of length $k$ and $n - k$ where $1 \leq k \leq n - 1$, and so an area of $k(n - k)$.) If the range of summation is extended to $\mathbb{Z}$, the change is of the order of $O(q^{-n^2/4})$. Convex polygons can be constructed by removing corner sites from these rectangles while preserving the perimeter. These "corners" are described by Ferrers diagrams, whose area–generating function is

$$F(q) = \frac{1}{(q;q)_\infty} = \prod_{k=1}^{\infty} \frac{1}{1-q^k} \,, \tag{1.9}$$

which is convergent for $|q| < 1$. A removal of one corner (ignoring overlaps) corresponds to multiplication with this area–generating function with the area weight replaced by $q^{-1}$. Correspondingly, the simultaneous removal of four corners corresponds to multiplication with $F(q^{-1})^4$, leading directly to the expression in (1.6).

The rest of the paper is set out as follows: in section 2 we state the two main theorems, where the first theorem compares the asymptotics of the finite–perimeter partition functions of all polygons with those of convex polygons while the second gives the asymptotics of various kinds of convex polygons, and our main result precisely, which combines these theorems to give the finite–perimeter partition function asymptotics for all polygons. In the following section 3 we prove the two main theorems. We end with a discussion of our results, including the derivation of the asymptotics as $q \to 1^+$ of the dominant asymptotic part (of the right–hand side) of (1.6).

## 2 Asymptotic Results

**Theorem 2.1** *Let $Z_n(q)$ and $Z_n^c(q)$ be the finite-perimeter partition functions of polygons and convex polygons, respectively, on the square lattice. Then, $Z_n(q) \sim Z_n^c(q)$ "exponentially fast" as $n \to \infty$: more precisely, for all $q > 1$ there exist $C > 0$ and $0 < \rho < 1$ such that for all integers $n > 1$*

$$1 \leq \frac{Z_n(q)}{Z_n^c(q)} < 1 + C\rho^n \,. \tag{2.1}$$

**Theorem 2.2** Let $Z_n^{(s)}(q)$ be the finite-perimeter partition function of rectangles ($s = 0$), Ferrers diagrams ($s = 1$), stacks or staircase polygons ($s = 2$), directed convex polygons ($s = 3$), and convex polygons ($s = 4$) on the square lattice. Then

$$Z_n^{(s)}(q) \sim Z_n^{(s),as}(q) \stackrel{def}{=} \frac{1}{(q^{-1};q^{-1})_\infty^s} \sum_{k=-\infty}^{\infty} q^{k(n-k)} \qquad (2.2)$$

exponentially fast as $n \to \infty$: more precisely, for all $q > 1$ there exist $C > 0$ and $0 < \rho < 1$ such that for all integers $n > 1$

$$1 - C\rho^n < \frac{Z_n^{(s)}(q)}{Z_n^{(s),as}(q)} < 1 \ . \qquad (2.3)$$

Note that we are using the symbol $Z_n^{(2)}(q)$, for ease of notation, to refer to either the finite–perimeter partition function of stacks or that of staircase polygons which are different functions. However, their dominant asymptotics in the case described above are identical.

The main result of our work is the following corollary, which follows directly from Theorems 2.1 and 2.2:

**Corollary 2.3** Let $Z_n(q)$ be the finite–perimeter partition function of polygons on the square lattice. Then

$$Z_n(q) \sim Z_n^{as}(q) \stackrel{def}{=} \frac{1}{(q^{-1};q^{-1})_\infty^4} \sum_{k=-\infty}^{\infty} q^{k(n-k)} = Z_n^{(4),as}(q) \qquad (2.4)$$

exponentially fast as $n \to \infty$: more precisely, for all $q > 1$ there exist $C > 0$ and $0 < \rho < 1$ such that for all integers $n > 1$

$$\left| \frac{Z_n(q)}{Z_n^{as}(q)} - 1 \right| < C\rho^n \ . \qquad (2.5)$$

*Proof of Corollary 2.3:* It follows from Theorem 2.1 and Theorem 2.2 (with $s = 4$) by multiplying the inequalities (2.1) and (2.3) that for $q > 1$ there exist $C > 0$ and $0 < \rho < 1$ such that

$$1 - C\rho^n < \frac{Z_n^c(q)}{Z_n^{as}(q)} \frac{Z_n(q)}{Z_n^c(q)} < 1 + C\rho^n \ . \qquad (2.6)$$

□

*Remark:* In the second theorem and the corollary the infinite sum could have been replaced by the finite–perimeter partition function of rectangles,

$$Z_n^{(0)} = \sum_{k=1}^{n-1} q^{k(n-k)} \ . \tag{2.7}$$

However, the form chosen has the advantage that one can write the $n$–dependence more explicitly:

$$Z_n^{as}(q) = \frac{q^{n^2/4}}{(q^{-1};q^{-1})_\infty^4} \begin{cases} \sum_{k=-\infty}^{\infty} q^{-k^2} & n \text{ even} \\ \sum_{k=-\infty}^{\infty} q^{-(k+1/2)^2} & n \text{ odd} \end{cases} \ . \tag{2.8}$$

## 3 Proofs of Theorems 2.1 and 2.2

In what follows, we denote the maximal area of a polygon with fixed perimeter $2n$ by $M(n)$. Clearly,

$$M(n) = \begin{cases} n^2/4 & n \text{ even} \\ (n^2-1)/4 & n \text{ odd} \end{cases} \ . \tag{3.1}$$

The proof of Theorem 2.1 will utilise two lemmata, the first one comparing polygons and nearly convex polygons, and the second one comparing nearly convex polygons with convex polygons. For this, we first define nearly convex more precisely.

**Definition 3.1** *A polygon on the square lattice is said to have* convexity index $\ell$, *if the difference between its perimeter and the perimeter of the bounding rectangle is equal to $2\ell$. For non–negative integer $\ell$ the set of* at-most-$\ell$-convex polygons *is defined to be the set of polygons with convexity index of at most $\ell$, and the corresponding finite–perimeter partition function is denoted by $Z_{n,\ell}^{ac}(q)$ (clearly, $Z_{n,0}^{ac}(q) = Z_n^c(q)$).*

**Lemma 3.2** *For all non–negative integers $\ell$ and for all $q$ such that $q^{\ell+1} > \mu_{saw}^4$ there exist $C > 0$ and $0 < \rho < 1$ such that for all integers $n > 1$*

$$1 \leq \frac{Z_n(q)}{Z_{n,\ell}^{ac}(q)} < 1 + C\rho^n \ , \tag{3.2}$$

where $\mu_{saw} \simeq 2.638$ is the connectivity constant of self–avoiding walks.

*Proof of Lemma 3.2:* The difference between the set of polygons and at-most-$\ell$-convex polygons is precisely the set of polygons with a convexity index of at least $\ell + 1$. These polygons have a bounding rectangle of half perimeter $\leq n - \ell - 1$, hence an area of at most $M(n - \ell - 1)$, and their number is clearly smaller than $c_n$, the total number of polygons with perimeter $2n$. Therefore we have the bound

$$0 \leq Z_n(q) - Z_{n,\ell}^{ac}(q) \leq c_n q^{M(n-\ell-1)} . \tag{3.3}$$

Rearranging terms and estimating $Z_{n,\ell}^{ac}(q) > q^{M(n)}$, this leads to

$$1 \leq \frac{Z_n(q)}{Z_{n,\ell}^{ac}(q)} \leq 1 + c_n q^{M(n-\ell-1)-M(n)} . \tag{3.4}$$

Now $c_n$ grows asymptotically as $\mu_{saw}^{2n}$ and we calculate

$$M(n - \ell - 1) - M(n) \leq -\frac{\ell + 1}{2}n + \frac{(\ell+1)^2 + 1}{4} . \tag{3.5}$$

Thus, provided that $q^{\frac{\ell+1}{2}} > \mu_{saw}^2$, we can find $C > 0$ and $0 < \rho < 1$ such that

$$c_n q^{M(n-\ell-1)-M(n)} < C\rho^n , \tag{3.6}$$

which completes the proof. $\square$

This lemma seems to suggest that the closer $q$ is to 1, the larger $\ell$ has to be chosen to get convergence. However, this is just an artefact of the rather simple estimation. One can sharpen the result with the help of the next lemma.

**Lemma 3.3** *For all non–negative integers $\ell$ and for all $q > 1$ there exist $C > 0$ and $0 < \rho < 1$ such that for all integers $n > 1$*

$$1 \leq \frac{Z_{n,\ell}^{ac}(q)}{Z_n^c(q)} < 1 + C\rho^n . \tag{3.7}$$

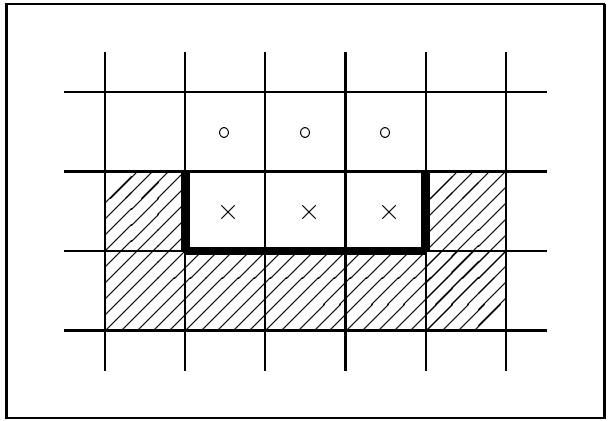

Figure 2: This figure shows the construction in Lemma 3.3. Shown is part of a polygon (shaded faces) with the thick line representing its border. The perimeter of the polygon is decreased by 2 and the convexity index decreased by 1 by adding the faces marked with × to the polygon. Note that the faces marked with ∘ are not part of the polygon, whereas the unmarked faces can be either.

*Proof of Lemma 3.3:* Let $Z'_{n,\ell}(q)$ denote the finite–perimeter partition function of polygons with convexity index $\ell$. Clearly $Z^{ac}_{n,\ell}(q) = \sum_{k=0}^{\ell} Z'_{n,k}(q)$. We first give an upper bound on $Z'_{n,\ell}(q)$ in terms of $Z'_{n-1,\ell-1}(q)$, valid for $\ell > 0$. To do this let us consider any polygon with perimeter $2n$ and convexity index $\ell$: we can add cells (faces of the lattice) to arrive at some polygon with perimeter $2(n-1)$ and convexity index $\ell - 1$ while preserving the bounding rectangle. As $\ell > 0$, we can always find an indentation within the polygon of the form depicted in Figure 2. Adding the marked faces to the polygon clearly changes perimeter and convexity index as desired. This implies that every polygon with perimeter $2n$ and convexity index $\ell$ can be constructed by removing cells (faces of the lattice) from a polygon with perimeter $2(n-1)$ and convexity index $\ell - 1$ while preserving the bounding rectangle.

By going through this procedure carefully, we will obtain the estimate

$$Z'_{n,\ell}(q) \leq \frac{2n}{q-1} Z'_{n-1,\ell-1}(q) \,. \tag{3.8}$$

To show this, we take any polygon with perimeter $2(n-1)$ and convexity index $\ell - 1$ and count the number of ways to remove faces. As the convexity index increases by exactly one, the faces to be removed have to be at the boundary of the polygon and have to be connected (one can of course get further such polygons by removing other sites that are not directly at the boundary, but then there is a smaller polygon with which we could have started the construction). There are less than $2n$ different faces of the polygon at the boundary. If we fix one face and start removing this one and additional faces in a clockwise order, we can remove only a finite number of faces, certainly less than $2n$. Each time we remove a face, the weight of the configuration gets reduced by $1/q$, and summing up the weights of all configurations generated in this way, we get a change of weight of at most $1/q + 1/q^2 + \ldots \leq 1/(q-1)$ by the removal of faces. Together with a multiplicity of at most $2n$ due to the choice of the first site, this implies the desired inequality (3.8).

Using this inequality, we get by iteration an upper bound for at-most-$\ell$-convex polygons in terms of convex polygons only:

$$Z^{ac}_{n,\ell}(q) \leq \sum_{k=0}^{\ell} \left(\frac{2n}{q-1}\right)^k Z^c_{n-k}(q) \,. \tag{3.9}$$

This leads to the need to estimate the terms in the sum on the right hand side of

$$1 \leq \frac{Z^{ac}_{n,\ell}(q)}{Z^c_n(q)} < 1 + \sum_{k=1}^{\ell} \left(\frac{2n}{q-1}\right)^k \frac{Z^c_{n-k}(q)}{Z^c_n(q)} \,. \tag{3.10}$$

With the help of the inequality

$$\frac{Z^c_n(q)}{Z^c_{n+1}(q)} \leq q^{-n/2} \,, \tag{3.11}$$

which follows from (3.16) in Lemma 3.4 with $s = 4$, we see now that each term of the sum in (3.10) is of the order of at most $n^\ell q^{-n/2}$. As $\ell$ is fixed, the sum contains only finitely many terms. Thus, if we pick a $\rho$ such that $q^{-1/2} < \rho < 1$ then there is a $C > 0$ such that

$$\sum_{k=1}^{\ell} \left(\frac{2n}{q-1}\right)^k \frac{Z^c_{n-k}(q)}{Z^c_n(q)} \leq C \rho^n \,, \tag{3.12}$$

which proves the lemma. □

Taken together, Lemma 3.2 and Lemma 3.3 prove Theorem 2.1.

*Proof of Theorem 2.1:* For any $q > 1$ we can choose $\ell$ fixed such that $q^{\ell+1} > \mu_{saw}^4$. Now we can write

$$1 \leq \frac{Z_n(q)}{Z_n^c(q)} = \frac{Z_n(q)}{Z_{n,\ell}^{ac}(q)} \frac{Z_{n,\ell}^{ac}(q)}{Z_n^c(q)} \leq (1 + C_1 \rho_1^n)(1 + C_2 \rho_2^n) , \qquad (3.13)$$

where the existence of $C_1 > 0$ and $0 < \rho_1 < 1$ is guaranteed by Lemma 3.2, and Lemma 3.3 guarantees the existence of $C_2 > 0$ and $0 < \rho_2 < 1$. It follows that for any $\max(\rho_1, \rho_2) < \rho < 1$ there exists a $C > 0$ such that

$$1 \leq \frac{Z_n(q)}{Z_n^c(q)} \leq 1 + C\rho^n . \qquad (3.14)$$

□

The inequality (3.11) used in the proof of Lemma 3.3 is contained in Lemma 3.4 (with $s = 4$), which we also use in a remark after the proof of Lemma 3.6.

**Lemma 3.4** *For $s \in \{0, 1, 2, 3, 4\}$ let $Z_n^{(s)}(q)$ be defined as in Theorem 2.2. Then, for any positive $q$ and integer $n > 1$ we have the inequality*

$$Z_{n+2}^{(s)}(q) \geq q^{n+1} Z_n^{(s)}(q) \qquad (3.15)$$

*and the slightly weaker bound*

$$Z_{n+1}^{(s)}(q) \geq q^{n/2} Z_n^{(s)}(q) . \qquad (3.16)$$

*Proof of Lemma 3.4:* If we increase the width of each row and then the height of each column of a convex polygon with perimeter $2n$ by one (by adding cells appropriately), we increase the perimeter by 4 and the area by $n + 1$. This implies immediately the first inequality. For the second one we have to labour slightly harder. We partition the set of convex polygons with respect to their bounding rectangles. Let $c_{(k,\ell)}^{m,(s)}$ denote the number of convex polygons of class $s$ with width $k$, height $\ell$, and area $m$. Then, by simply increasing the width or height of each row or column, respectively, of a polygon by one, we get the estimates

$$c^{m,(s)}_{(k+1,\ell)} \geq c^{m-\ell,(s)}_{(k,\ell)} \quad \text{and} \quad c^{m,(s)}_{(k,\ell+1)} \geq c^{m-k,(s)}_{(k,\ell)} \ . \tag{3.17}$$

(We need to treat both cases, as stacks ($s = 2$) lack reflection symmetry.) If we define the partition function $Z^{(s)}_{(k,\ell)}(q) = \sum_m q^m c^{m,(s)}_{(k,\ell)}$ then this implies the inequalities

$$Z^{(s)}_{(k+1,\ell)}(q) \geq q^\ell Z^{(s)}_{(k,\ell)}(q) \quad \text{and} \quad Z^{(s)}_{(k,\ell+1)}(q) \geq q^k Z^{(s)}_{(k,\ell)}(q) \ . \tag{3.18}$$

As $Z^{(s)}_{n+1}(q) = \sum_{k=0}^{n-1} Z^{(s)}_{(k+1,n-k)}$, we can now estimate

$$Z^{(s)}_{n+1}(q) \geq Z_{(1,n)}(q) + \sum_{k=1}^{n-1} q^{n-k} Z^{(s)}_{(k,n-k)}(q) \tag{3.19}$$

and

$$Z^{(s)}_{n+1}(q) \geq Z_{(n,1)}(q) + \sum_{k=1}^{n-1} q^k Z^{(s)}_{(k,n-k)}(q) \ , \tag{3.20}$$

whence it follows that

$$Z^{(s)}_{n+1}(q) \geq \sum_{k=1}^{n-1} \frac{q^{n-k} + q^k}{2} Z^{(s)}_{(k,n-k)}(q) \geq q^{n/2} Z^{(s)}_n(q) \ , \tag{3.21}$$

where we have used the geometric–arithmetic mean inequality. □

A simple idea of over–counting gives the upper bound for the partition function $Z^{(s)}_n(q)$ in the next lemma.

**Lemma 3.5** *For $s \in \{0, 1, 2, 3, 4\}$ let $Z^{(s)}_n(q)$ be defined as in Theorem 2.2. Then for any $q > 1$ and integer $n > 1$ we have the bound*

$$Z^{(s)}_n(q) < Z^{(s),as}_n(q) = \frac{1}{(q^{-1}; q^{-1})^s_\infty} \sum_{k=-\infty}^{\infty} q^{k(n-k)} \ . \tag{3.22}$$

*Proof of Lemma 3.5:* Every configuration in these models can be constructed by removing $s$ Ferrers diagrams from specified corners of rectangles with the restriction that the resulting configuration is still a polygon (this procedure does not change the perimeter). If one removes this restriction, one clearly over–counts. As the removal of Ferrers diagrams of arbitrary size is equivalent to multiplying the weight of the rectangle with $(q^{-1};q^{-1})_\infty^{-1}$, this implies for the generating function the inequality

$$Z_n^{(s)}(q) \leq \frac{Z_n^{(0)}(q)}{(q^{-1};q^{-1})_\infty^s} \ . \tag{3.23}$$

Replacing $Z_n^{(0)}(q) = \sum_{k=1}^{n-1} q^{k(n-k)}$ by the infinite sum $\sum_{k=-\infty}^{\infty} q^{k(n-k)}$ proves the lemma.

□

As a consequence of Lemma 3.4 and Lemma 3.5 we can now establish the desired convergence to $Z_n^{(s),as}(q)$. This is done in Lemma 3.6 in which we also establish the rate of convergence.

**Lemma 3.6** *For $s \in \{0,1,2,3,4\}$ let $Z_n^{(s)}(q)$ be defined as in Theorem 2.2. Then for all $q > 1$ there exist $C > 0$ and $0 < \rho < 1$ such that for all integers $n > 1$*

$$q^{-M(n)} \left( Z_n^{(s),as}(q) - Z_n^{(s)}(q) \right) < C\rho^n \ . \tag{3.24}$$

*Proof of Lemma 3.6:* We first consider $Z_{2n}^{(s)}(q)/q^{M(n)}$ and $Z_n^{(s),as}(q)/q^{M(n)}$ as series in $q^{-1}$ and show that we have convergence for each of the series coefficients. In order to compare the coefficients, we need to look more closely at the error made by the over–counting procedure. The over–counting results from Ferrers diagrams that touch each other, respectively from Ferrers diagrams that do not fit into the rectangle. In either case, this necessitates a minimal area removal of size $\min(k, n-k)$ from a $k \times (n-k)$–rectangle. Thus, the maximal weight of the excess configurations is

$$q^{k(n-k)-\min(k,n-k)} \ . \tag{3.25}$$

As both $Z_n^{(s)}(q)$ and $Z_n^{(0)}(q)/(q^{-1};q^{-1})_\infty^s$ have a leading power of $q^{M(n)}$, this implies that they agree in their leading $\lfloor \frac{n}{2} \rfloor$ coefficients, if considered as a series in $q^{-1}$.

If we define for $k = 0, 1, 2, \ldots$ the positive numbers

$$d_k^{(s),even} = [q^{-k}] \frac{1}{(q^{-1}; q^{-1})_\infty^s} \sum_{\ell=-\infty}^{\infty} q^{-\ell^2} \qquad (3.26)$$

$$d_k^{(s),odd} = [q^{-k}] \frac{1}{(q^{-1}; q^{-1})_\infty^s} \sum_{\ell=-\infty}^{\infty} q^{-\ell(\ell+1)} \qquad (3.27)$$

where $[q^{-k}]$ denotes the $k$-th coefficient of $Z_n^{(s),as}(q)/q^{M(n)}$ in $q^{-1}$, then these coefficients coincide with those of $Z_{2n}^{(s)}(q)/q^{M(n)}$, as explained above, for the first $n$ terms. This coincidence and the upper bound in Lemma 3.5 imply that

$$\sum_{k=0}^{n} q^{-k} d_k^{(s),even} \leq Z_{2n}^{(s)}(q)/q^{n^2} \leq \sum_{k=0}^{\infty} q^{-k} d_k^{(s),even} \qquad (3.28)$$

$$\sum_{k=0}^{n} q^{-k} d_k^{(s),odd} \leq Z_{2n+1}^{(s)}(q)/q^{n(n+1)} \leq \sum_{k=0}^{\infty} q^{-k} d_k^{(s),odd}, \qquad (3.29)$$

which in turn imply that the error is less than the error made by truncating the expansion of the upper bound in $q^{-1}$ after $n$ terms. As the left–hand sides converge exponentially fast in $q^{-1}$ to the right–hand sides, we can now write down the rate of convergence for the middle terms. More precisely, we have shown that for $0 < \rho < 1$ there exists a $C > 0$ such that for all $q > \rho^{-1}$

$$\frac{1}{(q^{-1}; q^{-1})_\infty^s} \sum_{k=-\infty}^{\infty} q^{-k^2} - \frac{1}{q^{n^2}} Z_{2n}^{(s)}(q) \leq C\rho^n \qquad (3.30)$$

$$\frac{1}{(q^{-1}; q^{-1})_\infty^s} \sum_{k=-\infty}^{\infty} q^{-k(k+1)} - \frac{1}{q^{n(n+1)}} Z_{2n+1}^{(s)}(q) \leq C\rho^n, \qquad (3.31)$$

which implies that for $0 < \rho < 1$ there exists a $C > 0$ such that for all $q > \rho^{-2}$

$$q^{-M(n)} \left( \frac{1}{(q^{-1}; q^{-1})_\infty^s} \sum_{k=-\infty}^{\infty} q^{k(n-k)} - Z_n^{(s)}(q) \right) < C\rho^n, \qquad (3.32)$$

which proves the lemma. □

*Remark:* By Lemma 3.4, we have the inequality

$$Z_{n+2}^{(s)}(q)/q^{(n+2)^2/4} \geq Z_n^{(s)}(q)/q^{n^2/4} \qquad (3.33)$$

which implies that the sequences $(Z_{2n}^{(s)}(q)/q^{n^2})$ and $(Z_{2n+1}^{(s)}(q))/q^{(n+1/2)^2})$ are monotonically increasing. Rewriting the upper bound of Lemma 3.5 gives the $n$–independent upper bounds

$$Z_n^{(s)}(q)/q^{n^2/4} < \frac{1}{(q^{-1};q^{-1})_\infty^s} \begin{cases} \sum_{k=-\infty}^\infty q^{-k^2} & n \text{ even} \\ \sum_{k=-\infty}^\infty q^{-(k+1/2)^2} & n \text{ odd} \end{cases}. \qquad (3.34)$$

Thus, the sequences $(Z_{2n}^{(s)}(q)/q^{n^2})$ and $(Z_{2n+1}^{(s)}(q))/q^{(n+1/2)^2})$ converge. One may be tempted to use this convergence and the fact that the series coefficients of $Z_{2n}^{(s)}(q)/q^{M(n)}$ and $Z_n^{(s),as}(q)/q^{M(n)}$ coincide for the leading $\lfloor \frac{n}{2} \rfloor$, as shown in the first part of the proof of Lemma 3.6, to show the sequences, $Z_{2n}^{(s)}(q)/q^{M(n)}$ and $Z_n^{(s),as}(q)/q^{M(n)}$, converge to the same (odd and even) limits. However, to use this convergence of the formal power series, and the point–wise convergence of the sequences, to imply equality of the limits one needs to utilise the positivity of the coefficients of the power series. This is precisely what was accomplished in the second part of the proof of Lemma 3.6, which also allowed us to estimate the rate of convergence simultaneously.

*Proof of Theorem 2.2:* This follows now directly from Lemma 3.6. □

## 4 Discussion

In this paper we have derived the leading asymptotic behaviour of the finite–perimeter generating function for polygons on the square lattice for area fugacity larger than one and have given a combinatorial interpretation of the result.

We conclude this paper by considering the behaviour of the form (1.6) when $q \to 1^+$. This is clearly far from being enough to determine the asymptotic behaviour of $Z_n(1)$, as one may not interchange the limits $n \to \infty$ and $q \to 1$.

We define

$$A_e(q) = \frac{\sum_{k=-\infty}^\infty q^{-k^2}}{(q^{-1};q^{-1})_\infty^4} \qquad (4.1)$$

and
$$A_o(q) = \frac{\sum_{k=-\infty}^{\infty} q^{-(k+1/2)^2}}{(q^{-1}; q^{-1})_\infty^4} . \tag{4.2}$$

Hence we can write

$$Z_n^{as}(q) = A(q)\, q^{n^2/4} , \tag{4.3}$$

where $A(q) = A_o(q)$ or $A(q) = A_e(q)$ when $n$ is restricted to subsequences with $n$ being odd or even respectively. The numerators of the functions $A_e(q)$ and $A_o(q)$ can be identified as limiting cases of the elliptic theta functions [18], that is,

$$\vartheta_3(0, q^{-1}) = \sum_{k=-\infty}^{\infty} q^{-k^2} \tag{4.4}$$

and

$$\vartheta_2(0, q^{-1}) = \sum_{k=-\infty}^{\infty} q^{-(k+1/2)^2} . \tag{4.5}$$

This allows the powerful theory of theta functions [18] to be utilised. In particular, the conjugate modulus transformation relates the theta functions of nome $p = e^{-\pi\eta} = q^{-1} < 1$ to theta functions of nome $p' = e^{-\pi/\eta}$. This is useful if we consider the asymptotics as $p \to 1^-$ (that is, $q \to 1^+$) since then $p' \to 0^+$. The conjugate modulus transformation yields

$$\vartheta_3(0, p) = \eta^{-1/2} \vartheta_3(0, p') \tag{4.6}$$

and

$$\vartheta_2(0, p) = \eta^{-1/2} \vartheta_4(0, p') = \eta^{-1/2} \sum_{k=-\infty}^{\infty} (-1)^k (p')^{k^2} . \tag{4.7}$$

Since

$$\vartheta_3(0, p') \sim \vartheta_4(0, p') \sim 1 \tag{4.8}$$

as $p' \to 0$, and since further [10]

$$(p; p)_\infty \sim \left(\frac{2}{\eta}\right)^{1/2} \exp\left[\frac{-\pi}{6\eta}\right] \tag{4.9}$$

as $p \to 1^-$ ($\eta \to 0^+$), the asymptotics of the functions $A_e(q)$ and $A_o(q)$ follow after some algebra. We hence obtain

$$A_e(q) \sim A_o(q) \sim \frac{1}{4}\left(\frac{\varepsilon}{\pi}\right)^{3/2} e^{2\pi^2/3\varepsilon} \qquad \text{as } q \to 1^+ , \tag{4.10}$$

where $\varepsilon = \log(q)$.

Lastly, we consider exact enumeration data for these models. Comparing

$$Z_n(q) / \sum_{k=-\infty}^{\infty} q^{k(n-k)} = \sum_{k=0}^{\infty} a_{n,k} q^{-k} \tag{4.11}$$

and

$$\frac{1}{(q^{-1};q^{-1})_\infty^4} = \sum_{k=0}^{\infty} b_k q^{-k} \tag{4.12}$$

we observe that the coefficients $a_{n,k}$ are monotonically increasing in $n$ and bounded above by $b_k$ for $n \leq 21$. Hence, we are led to conjecture that $Z_n^{as}(q)$ from (2.4) may, in fact, be a strict upper bound for $Z_n(q)$. We leave this as an open question.

## Acknowledgements

Financial support from the Australian Research Council is gratefully acknowledged by A.L.O. while T.P. thanks the Department of Mathematics at the University of Oslo and the Department of Physics at the University of Manchester, both where parts of this work were completed. This work was supported by EC Grant ERBCHBGCT939319 of the "Human Capital and Mobility Program" and EPSRC Grant No. GR/K79307. The authors thank the referees for their careful comments on our work.